\documentclass[a4paper,fleqn]{cas-dc}
\usepackage{amsmath,amssymb,amsfonts}
\usepackage{algorithmic}
\usepackage{graphicx}
\usepackage{textcomp}
\def\BibTeX{{\rm B\kern-.05em{\sc i\kern-.025em b}\kern-.08em
    T\kern-.1667em\lower.7ex\hbox{E}\kern-.125emX}}

\usepackage[utf8]{inputenc}
\usepackage[authoryear,longnamesfirst]{natbib}

\def\tsc#1{\csdef{#1}{\textsc{\lowercase{#1}}\xspace}}
\tsc{WGM}
\tsc{QE}

\usepackage{lineno,hyperref,amsfonts,amsmath,braket}
\usepackage{graphicx}
\usepackage{epstopdf}
\usepackage{algorithmic}
\usepackage{lipsum}
\usepackage{amsopn}
\usepackage[noabbrev]{cleveref}
\usepackage{color,soul}
\modulolinenumbers[5]

\newcommand{\e}{\varepsilon}

\newcommand{\tr}{\operatorname{tr\,}}

\newcommand{\Ree}{\operatorname{Re\,}}

\newcommand{\T}{\operatorname{T\,}}
\newcommand{\ad}{\operatorname{ad\,}}

\newcommand{\id}{\operatorname{id\,}}

\newtheorem{theorem}{\hspace{\parindent}\bf Theorem}[section]

\begin{document}
\let\WriteBookmarks\relax
\def\floatpagepagefraction{1}
\def\textpagefraction{.001}

\title [mode = title]{Influence of the Commutator Properties of Hamiltonians on the Robustness of Quantum Circuits}  

\shorttitle{Influence of the Commutator Properties} 
\shortauthors{Slynko, Bivziuk} 
%\author{Vladyslav Bivziuk, Vitalii Slynko
%\thanks{This work was partially supported by the DFG grant number SL343/1-1. }
%\thanks{Vitalii Slynko is with the Institute of Mathematics, University of Würzburg, Germany (e-mail:  vitalii.slynko@mathematik.uni-wuerzburg.de).}
%\thanks{Vladyslav Bivziuk is with the Department of Mathemetics, University of Illinois Urbana-Champaign, Champaign, IL 61820, USA (e-mail: \mbox{bivziuk2@illinois.edu}).}}

\tnotemark[1] 

\tnotetext[1]{This work was partially supported by the DFG grant number SL343/1-1. }

\author[1]{Vitalii Slynko}[style=chinese,      auid=000,       bioid=1,      orcid=0000-0001-8231-0130]

\cormark[1]


\ead{vitalii.slynko@mathematik.uni-wuerzburg.de}

\credit{Methodology, work with manuscript}

\affiliation[1]{organization={Institute of Mathematics, University of Würzburg},
            city={Würzburg},
%          citysep={}, % Uncomment if no comma needed between city and postcode
            postcode={97074}, 
            state={Bayern},
            country={Germany}}

\author[b]{Vladyslav Bivziuk}

% Email id of the second author
\ead{bivziuk2@illinois.edu}

\credit{Work with manuscript, software}

\affiliation[b]{organization={Department of Mathemetics, University of Illinois Urbana-Champaign}, 
            city={Champaign},
%          citysep={}, % Uncomment if no comma needed between city and postcode
            postcode={61820}, 
            state={Illinois},
            country={USA}}

\cortext[1]{Corresponding author}

\begin{abstract}
We have proved new estimates for the coherent control errors of quantum circuits used in quantum computing. These estimates essentially take into account the commutator properties of the Hamiltonians and are based on the formulas of the commutator calculus.
\end{abstract}


\maketitle

\section{Introduction}

The rapid development of quantum computing in recent years poses many new engineering, physical and mathematical problems.
An obstacle to the successful practical implementation of quantum computing algorithms is the influence of a significant amount of noise, which is a problem for demonstrating quantum advantages. Typically, the noise arising in quantum devices is divided into two types: coherent and decoherent. From a mathematical point of view, a coherent error can be represented as self-adjoint perturbations of the system's Hamiltonian, and decoherent error can be represented as non-self-adjoint perturbations. Studies of coherent control errors were previously considered in \cite{5,7,8,9}.
 In a recent paper \cite{BFCh2023}, the relevance of studying estimates of coherent errors is substantiated in detail and some results of the robustness analysis of quantum algorithms to coherent control errors are presented. The main theoretical result is justified using the Lipschitz constants. The estimates for the fidelity bounds contain the norms of the Hamiltonians that generate individual circuit elements. The results are applied to the 3-qubit quantum Fourier transform. The essential difference of these results from the well-known \cite{7,8,9,28} is that the perturbations of the Hamiltonians are not assumed to be small, identical or subject to certain statistics as, for example, in \cite{28}.

% The rapid development of quantum computing in recent years poses many new engineering, physical and mathematical problems. One such problem is assessing the effect of noise on quantum circuits made up of reliable qubits and gates. The relevance and importance of this problem for the practical implementation of quantum computing devices is fully justified in a recent paper \cite{BFCh2023}.
% Estimates for coregent control errors were also obtained there based on the norms of the Hamiltonians of the elements of a quantum mechanical circuit. From a mathematical point of view, the coherent control error can be described as a small perturbation $\varepsilon \wt H=\varepsilon H$ of the unperturbed Hamiltonian $H$ of an ideal quantum gate under the condition that the operator $\wt H$ is self-adjoint $\wt H^{\dagger}=\wt H$. Rejection of the condition of self-adjointness of the perturbation $\wt H$ leads to incoherent errors, which will not be considered here.

The aim of this work is to develop new theoretical tools for the analysis of coherent errors in the same problem statement as in \cite{BFCh2023}. Our main idea is to take into account the commutative properties of Hamiltonians that generate an ideal quantum circuit. To do this, we use the mathematical apparatus of commutator calculus \cite{MKS}. We note that the commutator calculus was previously used for various problems of stability theory \cite{BSlT2019,ADSl2023} and control theory \cite{Zu,Sutner,Grush} as well as problems of quantum mechanics and quantum control \cite{jones,LEVITT,merill}. The main contribution of this work is that we show that taking into account the commutative properties of the Hamiltonians of a quantum circuit in some cases allows us to establish less conservative error estimates than those obtained in \cite{BFCh2023}. We obtain estimates for two types of coherent perturbations (single-parameter and multi-parameter) and present illustrative examples.

The work is organized as follows. In the next section, we give the necessary notation, facts, and commutator formulas. In Section 3, we formulate the problem, and in the fourth section, we prove the main result of the paper. In the fifth section, we give an illustrative examples and compare our results with known ones. In the last section, we discuss the results and prospects for further extensions.

\section{Notations and commutator formulas}
For the state vector $\ket{\psi}\in \mathbb C^n$, the norm $\|\ket{\psi}\|$ is the standard 2-norm, i.e., $\|\ket{\psi}\|:=\sqrt{\braket{\psi,\psi}}$, for the vector $x=(x_1,\dots,x_N)^{\T}\in\mathbb R^N$, the norm is $\|x\|_{\infty}:=\max_{k=\overline{1,N}}|x_k|$.

For an arbitrary matrix $A\in \Bbb C^{n\times n}$, its conjugate matrix is denoted by $A^{\dagger}$. The norm of $A$ is defined by $\|A\|_2:=\lambda_{\max}^{1/2}(A^{\dagger}A)$, i.e., the maximum singular value of $A$.

Let $A$, $B$ $\in \Bbb{C}^{n \times n}$. Then, 

\begin{equation*}
\gathered
[A,B]:= AB-BA, \quad \ad A\,:\,  \Bbb{C}^{n \times n} \to  \Bbb{C}^{n \times n},\\ \quad X\mapsto \ad A(X):=[A,X].
\endgathered
\end{equation*}
An important role is played by the Hadamard formula \cite{MKS}
\begin{equation}\label{0}
\gathered
e^{t \ad A}(B)=e^{tA}Be^{-tA}, \quad t \in \Bbb{R}
\endgathered
\end{equation}

 The group of unitary matrices is defined as
\begin{equation*}
\gathered
\Bbb{U}(n):=\{A \in  \Bbb{C}^{n \times n} \, :\, A^{\dagger}A=AA^{\dagger}=\id\}.
\endgathered
\end{equation*}

If $U \in \Bbb{U}(n)$, then there exists a self-adjoint matrix $A$ (generator of $U$), i.e., $A^{\dagger}=A$ such that $U=e^{-iA}$.  If $A^{\dagger}=A$, $U\in \Bbb{U}(n)$, then 

\begin{equation}\label{fact0}
\gathered
\|U\|_2=1, \quad \|e^{it \ad A}(B)\|_2=\|B\|_2.
\endgathered
\end{equation}
The identity matrix is denoted by $I$.

\section{Problem statement}

Let $\ket{\psi_0}$ be the initial state of the quantum mechanical system, which is assumed to be normalized $\|\ket{\psi_0}\|=1$. In ideal quantum circuits, the final state vector $\ket{\widehat{\psi}}$ is defined as the result of the sequential action of the unitary operators 
 $\widehat{U}_1, \dots, \widehat{U}_N$, $\widehat{U}_k\in \Bbb{U}(n)$, $k=\overline{1,N}$ acting on the initial state of the system  $\ket{\psi_0}$. For each unitary operator $\widehat{U}_k$, there exists a Hamiltonian $H_k$, $H_k^{\dagger}=H_k$ that generates $\widehat{U}_k=e^{-iH_k}$. 
 
Therefore, the final state is defined as
\begin{equation}\label{fact0}
\gathered
\ket{\widehat{\psi}}=e^{-iH_1}\dots e^{-iH_N}\ket{\psi_0}.
\endgathered
\end{equation}

 In a non-ideal quantum circuit, the final state can be described in various ways. We restrict ourselves to two variants of perturbed states. For $\varepsilon\in \Bbb{R}$, we introduce the perturbed state
\begin{equation}\label{1}
\gathered
\ket{\psi(\varepsilon)}:=U_1(\varepsilon)\dots U_N(\varepsilon)\ket{\psi_0},
\endgathered
\end{equation}
where $U_k(\varepsilon)=e^{-i(1+\varepsilon)H_k}$, $k=\overline{1,N}$. 

For $\epsilon=(\varepsilon_1,\dots, \varepsilon_N)^{\T}\in \Bbb{R}^N$, we introduce a perturbed state of the form
\begin{equation}\label{2}
\gathered
\ket{\psi(\epsilon)}:=U_1(\varepsilon_1)\dots U_N(\varepsilon_N)\ket{\psi_0}.
\endgathered
\end{equation}

The main mathematical problem that we are considering formulated in \cite{BFCh2023} is to find $M(\overline{\varepsilon})\geq 0$ for a given  $\overline{\varepsilon}\geq 0$ such that for $\varepsilon\in \Bbb{R}$ ($\epsilon \in \Bbb{R}^N$ for the state \eqref{2}) under the condition $|\varepsilon|\le\overline{\varepsilon}$ (respectively $\|\epsilon\|_{\infty}\leq \overline{\varepsilon}$) and any initial state $\ket{\psi_0}$, $\|\ket{\psi_0}\|_2=1$, the following estimate holds  
\begin{equation}\label{3}
\gathered
|\braket{\psi(\varepsilon),\widehat{\psi}}|\geq 1- M(\overline{\varepsilon})\quad 
\endgathered
\end{equation}
(respectively $|\braket{\psi(\epsilon),\widehat{\psi}}|\geq 1- M(\overline{\varepsilon})$).

The main result obtained in \cite{BFCh2023} is that $M(\overline{\varepsilon})$ can be chosen in the form
\begin{equation}\label{Jul}
\gathered
M(\overline{\varepsilon})=\frac{\overline{\varepsilon}^2}{2}\Big(\sum\limits_{k=1}^N\|H_k\|_2\Big)^2.
\endgathered
\end{equation}
The aim of our paper is to refine this estimate using the commutation properties of the set of Hamiltonians $H_k$, $k=\overline{1,n}$. We want to show that taking these properties into account can lead in some cases to less conservative estimates for $|\braket{\psi(\varepsilon),\widehat{\psi}}|$.

\section{Main result}

We introduce the notation
\begin{equation*}
\centering
\gathered
A_1:=H_1, \\ A_k:=e^{-iH_1}\dots e^{-iH_{k-1}}H_ke^{iH_{k-1}}\dots e^{iH_1},\\
\mathcal{H}_0:=A_1+\dots+A_N,\\
\eta_{2p}:=\|(\ad H_1)^p(H_2)\|_2, \\ \eta_{kp}:=\|(\ad H_{k-1})^p(H_k)\|_2+\\ \|(\ad H_{k-2})^p(e^{-iH_{k-1}}H_ke^{iH_{k-1}})\|_2+\dots\\
+ \|(\ad H_1)^p(e^{-iH_2}\dots e^{-iH_{k-1}}H_ke^{iH_{k-1}}\dots e^{iH_2})\|_2, \\ k=\overline{2,N}
\endgathered
\end{equation*}

\begin{theorem}\label{Thm1}
For the perturbed state \eqref{1}, the estimate \eqref{3} is satisfied with
\begin{equation}\label{4}
\gathered
M(\overline{\varepsilon})=\frac{1}{2}\Big(\|\mathcal{H}_0\|_2+\sum\limits_{k=2}^{N}\sum\limits_{p=1}^{\infty}\frac{\overline{\varepsilon}^p
\eta_{kp}}{(p+1)!}\Big)^2\overline{\varepsilon}^2.
\endgathered
\end{equation}
\end{theorem}

For the perturbed state \eqref{2}, the following statement is true.

\begin{theorem}\label{Thm2}
Let
\begin{equation*}
\gathered
h_0=\max\limits_{\lambda \in \Bbb{R}^N,\, \|\lambda\|_{\infty}=1}\|\lambda_1 A_1+\dots+\lambda_N A_N\|_2.
\endgathered
\end{equation*}
Then, the perturbed state \eqref{2} satisfies the estimate \eqref{3} with
\begin{equation}\label{5}
\gathered
M(\overline{\varepsilon})=\frac{1}{2}\Big(h_0+\sum\limits_{k=2}^{N}\sum\limits_{p=1}^{\infty}\frac{\overline{\varepsilon}^p\eta_{kp}}{(p+1)!}\Big)^2\overline{\varepsilon}^2.
\endgathered
\end{equation}
\end{theorem}

\section{Proof of the main result}

{\it Proof} of Theorem \ref{Thm1}.

Let 
\begin{equation}\label{6}
\gathered
\ket{\psi(t)}:=e^{-i(1+t)H_1} \dots e^{-i(1+t)H_N}\ket{\psi_0}.
\endgathered
\end{equation}
Differentiating with respect to $t$, we find
\begin{equation*}
\gathered
\ket{\dot{\psi}(t)}=-iH_1e^{-i(1+t)H_1}\dots e^{-i(1+t)H_N}\ket{\psi_0}-\dots \\
-ie^{-i(1+t)H_1}\dots e^{-i(1+t)H_{k-1}}H_k e^{-i(1+t)H_k}\dots e^{-i(1+t)H_N}\ket{\psi_0}\\
-ie^{-i(1+t)H_1}\dots e^{-i(1+t)H_N}H_N\ket{\psi_0}.
\endgathered
\end{equation*}
From the formula \eqref{0}, it follows that (for details see Appendix A)
\begin{equation*}
\gathered
e^{-i(1+t)H_1}H_2=e^{-i(1+t)\ad H_1}(H_2) e^{-i(1+t)H_1}.
\endgathered
\end{equation*}

\begin{equation*}
\gathered
e^{-i(1+t)H_1}\dots e^{-i(1+t)H_{k-1}}H_k\\
=e^{-i(1+t)\ad H_1}\dots e^{-i(1+t)\ad H_{k-1}}(H_k)\\
\times e^{-i(1+t)H_1}\dots e^{-i(1+t)H_{k-1}}.
\endgathered
\end{equation*}

Therefore, from \eqref{6} we get
\begin{equation}\label{7}
\gathered
\ket{\dot{\psi}(t)}=-i\mathcal{H}(t)\ket{\psi(t)},
\endgathered
\end{equation}
where 
\begin{equation*}
\gathered
\mathcal{H}(t)=H_1+e^{-i(1+t)\ad H_1}(H_2)+
\dots\\+ e^{-i(1+t)\ad H_{1}}\dots e^{-i(1+t)\ad H_{N-1}}(H_N).
\endgathered
\end{equation*}
We denote  $\delta\mathcal H(t):=\mathcal H(t)-\mathcal H_0$ and represent \eqref{7}  as
\begin{equation*}
\gathered
\ket{\dot{\psi}(t)}=-i(\mathcal{H}_0+\delta\mathcal{H}(t))\ket{\psi(t)},
\endgathered
\end{equation*}
Integrating from $0$ to $\varepsilon$, we get
\begin{equation*}
\gathered
\ket{\psi(\varepsilon)}=\ket{\widehat{\psi}}-i\int\limits_0^{\varepsilon}(\mathcal{H}_0+\delta\mathcal{H}(t))\ket{\psi(t)}\,dt.
\endgathered
\end{equation*}
Hence, we find the estimate
\begin{equation}\label{9}
\gathered
\|\ket{\psi(\varepsilon)}-\ket{\widehat{\psi}}\|\le\int\limits_0^{\overline{\varepsilon}}(\|\mathcal{H}_0\|_2+\|\delta\mathcal{H}(t)\|_2)\|\ket{\psi(t)}\|\,dt\\
=\int\limits_0^{\overline{\varepsilon}}(\|\mathcal{H}_0\|_2+\|\delta\mathcal{H}(t)\|_2)\,dt=\|\mathcal{H}_0\|_2\overline{\varepsilon}+
\int\limits_0^{\overline{\varepsilon}}\|\delta\mathcal{H}(t)\|_2\,dt.
\endgathered
\end{equation}
To estimate $\delta\mathcal H(t)$, we represent it in the form
\begin{equation*}
\gathered
\delta\mathcal H(t)=(e^{-i(1+t)\ad H_1}(H_2)-e^{-i\ad H_1}(H_2))\\
+\dots
+(e^{-i(1+t)\ad H_1}\dots e^{-i(1+t)\ad H_{N-1}}(H_N)\\- 
e^{-i\ad H_1}\dots e^{-i\ad H_{N-1}}(H_N)).
\endgathered
\end{equation*}
We estimate the norm of the expression in the first bracket
\begin{equation*}
\gathered
\|e^{-i(1+t)\ad H_1}(H_2)-e^{-i\ad H_1}(H_2)\|_2\\
=\|e^{-i\ad H_1}(e^{-it\ad H_1}(H_2)-H_2)\|_2 \\= 
\|e^{-it\ad H_1}(H_2)-H_2\|_2\\
\le
\sum\limits_{p=1}^{\infty}\frac{|t|^p}{p!}\|(\ad H_1)^p(H_2)\|_2=
\sum\limits_{p=1}^{\infty}\frac{|t|^p\eta_{2p}}{p!}.
\endgathered
\end{equation*}
It follows from the inequality \eqref{AppB} (see Appendix B) that
\begin{equation*}
\gathered
\|e^{-i(1+t)\ad H_1}\dots e^{-i(1+t)\ad H_{k-1}}(H_k)\\-
e^{-i\ad H_1}\dots e^{-i\ad H_{k-1}}(H_k)\|_2
\le \sum\limits_{p=1}^{\infty}\frac{|t|^p\eta_{kp}}{p!}.
\endgathered
\end{equation*}
Thus,
\begin{equation*}
\gathered
\int\limits_0^{\overline{\varepsilon}}\|\delta\mathcal H(t)\|_2\,dt\le\int\limits_0^{\overline{\varepsilon}}\sum\limits_{k=2}^{N}
\sum\limits_{p=1}^{\infty}\frac{|t|^p\eta_{kp}}{p!}\,dt=
\sum\limits_{k=2}^{N}
\sum\limits_{p=1}^{\infty}\frac{\overline{\varepsilon}^{p+1}\eta_{kp}}{(p+1)!}.
\endgathered
\end{equation*}
From \eqref{9}, we find
\begin{equation*}
\gathered
\|\ket{\psi(\varepsilon)}-\ket{\widehat{\psi}}\|\le
\Big(\|\mathcal H_0\|_2+\sum\limits_{k=2}^{N}\sum\limits_{p=1}^{\infty}\frac{\eta_{kp}\overline{\varepsilon}^p}{(p+1)!}\Big)\overline{\varepsilon}\\= \sqrt{2M(\overline{\varepsilon})}.
\endgathered
\end{equation*}
Therefore,
\begin{equation*}
\gathered
2M(\overline{\varepsilon})\ge \|\ket{\psi(\varepsilon)}-\ket{\widehat{\psi}}\|^2 \\=
2-2\Ree \braket{\psi(\varepsilon)|\widehat{\psi}}\ge
2-2|\braket{\psi(\varepsilon)|\widehat{\psi}}|,
\endgathered
\end{equation*}
which completes the proof of Theorem \ref{Thm1}. The theorem is proved.

{\it Proof} of Theorem \ref{Thm2}. The proof of this theorem is similar to the proof of the previous one; therefore, we indicate only its essential changes. We recall that $\ket{\psi(\epsilon)}$, $\epsilon\in\Bbb R^N$ is defined from \eqref{2}. We denote $\ket{\Psi(t)}:=\ket{\psi(t\epsilon)}$. Then,
\begin{equation}\label{9*}
\gathered
\ket{\dot{\Psi}(t)}=-i\mathcal{H}(\epsilon,t)\ket{\Psi(t)},
\endgathered
\end{equation}
where
\begin{equation*}
\gathered
\mathcal{H}(\epsilon,t)=\varepsilon_1H_1+\varepsilon_2e^{-i(1+\varepsilon_1t)\ad H_1}(H_2)+
\dots\\
+\varepsilon_Ne^{-i(1+\varepsilon_1t)\ad H_{1}}\dots e^{-i(1+\varepsilon_{N-1}t)\ad H_{N-1}}(H_N).
\endgathered
\end{equation*}
We denote $\delta\mathcal H(\epsilon,t):=\mathcal H(\epsilon,t)-\mathcal H(\epsilon,0)$.
Then,
\begin{equation*}
\gathered
\mathcal H(\epsilon,0)=\varepsilon_1 A_1+\dots+\varepsilon_N A_N,\\
\delta\mathcal H(\epsilon,t)=
\varepsilon_2\big(e^{-i(1+\varepsilon_1t)\ad H_1}(H_2)-e^{-i\ad H_1}(H_2)\big)+\dots\\
+\varepsilon_N\big(e^{-i(1+\varepsilon_1t)\ad H_1}\dots e^{-i(1+\varepsilon_{N-1}t)\ad H_{N-1}}(H_N) \\-
e^{-i\ad H_1}\dots e^{-i\ad H_{N-1}}(H_N)\big)
\endgathered
\end{equation*}
It follows from the definition of the constant $h_0$ that
\begin{equation*}
\gathered
\|\mathcal H_0(\epsilon,0)\|_2\le
\max\limits_{\lambda\in\Bbb R^N,\,\|\lambda\|_{\infty}\le 1}
\|\lambda_1A_1+\dots+\lambda_NA_N\|_2
\overline{\varepsilon}\\
=\max\limits_{\lambda\in\Bbb R^N,\,\|\lambda\|_{\infty}= 1}
\|\lambda_1A_1+\dots+\lambda_NA_N\|_2
\overline{\varepsilon}=
h_0\overline{\varepsilon},\\ \lambda=(\lambda_1,\dots,\lambda_N).
\endgathered
\end{equation*}
Taking into account the estimates \eqref{AppB*} and \eqref{AppB} (see Appendix B), we obtain the inequality
\begin{equation*}
\gathered
\|\delta\mathcal H(\epsilon,t)\|_2\le \overline{\varepsilon}\sum_{k=2}^{N}\sum\limits_{p=1}^{\infty}
\frac{(|t|\overline{\varepsilon})^{p}\eta_{kp}}{p!}.
\endgathered
\end{equation*}
The rest of the proof does not differ from the proof of Theorem \ref{Thm1}. The theorem is proved.

{\it Remark.} The main difficulty in applying Theorem \ref{Thm2} to the perturbed state \eqref{2} is the calculation of the constant $h_0$ which is reduced to solving the optimization problem
\begin{equation*}
\gathered
\max\limits_{\lambda\in\Bbb R^N,\,\|\lambda\|_{\infty}= 1}
\|\lambda_1A_1+\dots+\lambda_NA_N\|_2.
\endgathered
\end{equation*}
For $N=2$ this problem can be solved by elementary means regardless of the dimensions of the matrices (see Appendix C). For the case of an arbitrary number of matrices, the solution to this problem is comprehensible to modern mathematical software, at least for matrices of small dimension. For high-dimensional matrices, this optimization problem is possibly an open problem requiring further investigation.

\section{Example and comparison of results.}
 {\it Example 1.} Consider the following simple example, borrowed from \cite{BFCh2023}, which shows the motivation for our approach. We apply a sequence of rotations
\begin{equation*}
U_{\text{ideal}}=R_z(\theta_z)R_y(\theta_y)
\end{equation*}
to the initial state $\psi_0$, where $R_z(\theta_z)$ and $R_y(\theta_y)$ denote the rotation around the $z$ and $y$ axes by the angles $\theta_z$ and $\theta_y$, respectively. We will assume that $\theta_z>0$ and $\theta_y>0$. We also consider the perturbed action
\begin{equation*}
U_{\text{noisy}}(\epsilon)=R_z(\theta_z(1+\varepsilon_z))R_y(\theta_y(1+\varepsilon_y)),
\end{equation*}
where $\epsilon=(\varepsilon_y,\varepsilon_z)\in\Bbb R^2$ are unknown noise parameters. We estimate the error $|\langle\psi(\epsilon),\widehat{\psi}\rangle|$, where $\psi(\epsilon)=U_{\text{noisy}}(\epsilon)\psi_0$,
$\widehat{\psi}=U_{\text{ideal}}\psi_0$ in two ways: using the result from \cite{BFCh2023} (see the \eqref{Jul} formula) and Theorem \ref{Thm2}.

We recall that the Pauli matrices
\begin{equation*}
X=
\begin{pmatrix}
0&1\\
1&0
\end{pmatrix},\quad
Y=i
\begin{pmatrix}
0&-1\\
1&0
\end{pmatrix},\quad
Z=\begin{pmatrix}
1&0\\
0&-1
\end{pmatrix}
\end{equation*}
allow expressing the rotations $R_z(\theta_z)$ and $R_y(\theta_y)$ explicitly
\begin{equation*}
R_z(\theta_z)=e^{-i\frac{1}{2}\theta_z Z},\quad R_y(\theta_y)=e^{-i\frac{1}{2}\theta_y Y}.
\end{equation*}
We recall properties of Pauli matrices 
\begin{equation}\label{Ex1}
[X,Y]=2iZ,\quad [Y,Z]=2iX,\quad [Z,X]=2iY.
\end{equation}
\begin{equation}\label{Ex2}
X^2=Y^2=Z^2=I, \quad YZY=-Z
\end{equation}
that will be useful for further calculations as well as the Euler's formula
\begin{equation}\label{Ex3}
e^{i\varphi X}=\cos\varphi I+i\sin\varphi X, 
\end{equation}
which remains valid when the matrix $X$ is replaced by the matrix $Z$ or $Y$.

To compare the results of error estimation obtained in Theorem \ref{Thm2} and in \cite{BFCh2023} (formula \eqref{Jul}), it is necessary to calculate the values
\begin{equation}\label{Ex4}
M_{\eqref{Jul}}(\overline{\varepsilon})=\frac{\overline{\varepsilon}^2}{2}(\|H_1\|_2+\|H_2\|_2)^2
\end{equation}
and
\begin{equation}\label{Ex5}
M_{\eqref{5}}(\overline{\varepsilon})=\frac{\overline{\varepsilon}^2}{2}\Big(h_0+\sum\limits_{p=1}^{\infty}\frac{\overline{\varepsilon}^p\eta_{2p}}{(p+1)!}\Big)^2.
\end{equation}
Here, $H_1=\frac{i\theta_y}{2}Y$, $H_2=\frac{i\theta_z}{2}Z$,
\begin{equation}
\gathered
h_0=\max\limits_{\lambda\in\Bbb R^2,\|\lambda\|_{\infty}=1}\|\lambda_1A_1+\lambda_1A_2\|_2,\\
A_1=H_1,\quad A_2=e^{-\frac{i\theta_y}{2}Y}H_2e^{\frac{i\theta_y}{2}Y}
\endgathered
\end{equation}
and $\eta_{2p}=\|(\ad H_1)^p(H_2)\|_2$, $p\in\Bbb N$.

First of all, we calculate the matrix $A_2$ using the Euler formula \eqref{Ex3} and the commutation relations for the Pauli matrices \eqref{Ex1} and the formula \eqref{Ex2}:
\begin{equation*}
\gathered
A_2=\frac{i\theta_z}{2}e^{-\frac{i\theta_y}{2}Y}Ze^{\frac{i\theta_y}{2}Y}\\
=\frac{i\theta_z}{2}(\cos\frac{\theta_y}{2}I-i\sin\frac{\theta_y}{2}Y)Z(\cos\frac{\theta_y}{2}I+i\sin\frac{\theta_y}{2}Y)\\
=\frac{i\theta_z}{2}(\cos^2\frac{\theta_y}{2}Z+\sin^2\frac{\theta_y}{2}YZY+i\sin\frac{\theta_y}{2}\cos\frac{\theta_y}{2}[Z,Y])\\
=\frac{i\theta_z}{2}(\cos\theta_y Z+\sin\theta_y X).
\endgathered
\end{equation*}
Consequently,
\begin{equation*}
\gathered
h_0=\max\limits_{\lambda\in\Bbb R^2,\|\lambda\|_{\infty}=1}\Big\|\frac{\lambda_2\theta_z}{2}\sin\theta_y X+\frac{\lambda_1\theta_y}{2}Y+\frac{\lambda_2\theta_z}{2}\cos\theta_y Z\Big\|_2\\=
\max\limits_{\lambda\in\Bbb R^2,\|\lambda\|_{\infty}=1}\frac{1}{2}\sqrt{\lambda_1^2\theta_y^2+\lambda_2^2\theta_z^2}
=\frac{1}{2}\sqrt{\theta_y^2+\theta_z^2}.
\endgathered
\end{equation*}
To calculate the constants $\eta_{2p}$, we note that
\begin{equation*}
\gathered
\eta_{2p}=\|(\ad H_1)^p(H_2)\|_2=\frac{\theta_y\theta_z^p}{2^{p+1}}\|(\ad Y)^p(Z)\|_2.
\endgathered
\end{equation*}
The commutation relations for the Pauli matrix \eqref{Ex1} imply that
$\|(\ad Y)^p(Z)\|_2=2^p$. Therefore, from \eqref{Ex4} and \eqref{Ex5}, we find  
\begin{equation*}
M_{\eqref{5}}(\overline{\varepsilon})=\frac{\overline{\varepsilon}^2}{2}\Big(\frac{1}{2}\sqrt{\theta_y^2+\theta_z^2}+
\frac{\theta_y}{2\theta_z\overline{\varepsilon}}(e^{\theta_z\overline{\varepsilon}}-1-\theta_z\overline{\varepsilon})\Big)^2.
\end{equation*}
\begin{equation*}
M_{\eqref{Jul}}(\overline{\varepsilon})=\frac{\overline{\varepsilon}^2}{8}(\theta_y+\theta_z)^2.
\end{equation*}
Theorem \ref{Thm2} leads to a better estimate for errors if the inequality
$M_{\eqref{5}}(\overline{\varepsilon})<M_{\eqref{Jul}}(\overline{\varepsilon})$ is satisfied. This inequality holds for all $\overline{\varepsilon}\in(0,\overline{\varepsilon}_{\max})$, where
$\overline{\varepsilon}_{\max}$ is the only positive root of the equation
\begin{equation*}
\frac{\theta_y}{\theta_z}(e^{\theta_z\overline{\varepsilon}_{\max}}-1-\theta_z\overline{\varepsilon}_{\max})=\overline{\varepsilon}_{\max}
(\theta_y+\theta_z-\sqrt{\theta_y^2+\theta_z^2}).
\end{equation*}
For example, for angles
$\theta_y=\frac{\pi}{2}$, $\theta_z=\frac{\pi}{4}$, we get $\overline{\varepsilon}_{\max}\approx 0.759$,
$M_{\eqref{5}}(\overline{\varepsilon}_{\max})=M_{\eqref{Jul}}(\overline{\varepsilon}_{\max})\approx 0.39819$.
For parameter values $\overline{\varepsilon}\in(0,0.759)$, Theorem \ref{Thm2} gives a better estimate for the error than the main result from \cite{BFCh2023}. For example, for $\overline{\varepsilon}=0.2$, Theorem \ref{Thm2} leads to the estimate
$|\langle\psi(\epsilon),\widehat{\psi}\rangle|\ge 0.9822095$,
while the main result from \cite{BFCh2023} leads to the estimate
$|\langle\psi(\epsilon),\widehat{\psi}\rangle|\ge 0.97224174$.
We note that the estimate obtained in the numerical experiment in \cite{BFCh2023} is $|\langle\psi(\epsilon),\widehat{\psi}\rangle|\approx 0.985$ which is close to our result.
 
{\it Example 2.} Let $\overline{\varepsilon}=0.1$ and the Hamiltonian matrices $H_1$ and $H_2$ be given by
\begin{equation*}
\gathered
H_1=\begin{pmatrix}
5&0.2\\
 0.2&-0.5
\end{pmatrix},\\
H_2=\begin{pmatrix}
 -0.48186&0.08813-3.29597i\\
 0.08813+3.29597i&4.99186
\end{pmatrix}.
\endgathered
\end{equation*}
 In this case, for the perturbed states \eqref{1} and \eqref{2}, the main result from \cite{BFCh2023} (see \eqref{Jul}) leads to the estimate $|\braket{\psi(\varepsilon),\widehat{\psi}}|\ge 0.49855$. For the pertubed state \eqref{1}, Theorem \ref{Thm1} leads to the estimate $|\braket{\psi(\varepsilon),\widehat{\psi}}|\ge 0.87333$. For the pertubed state \eqref{2}, Theorem \ref{Thm2} leads to a less conservative estimate $|\braket{\psi(\varepsilon),\widehat{\psi}}|\ge 0.84133$ than the main result from \cite{BFCh2023}.

\section{Discussion of results} 

We note that from the triangle inequality it follows that for all $\lambda\in\Bbb R^N$, $\|\lambda\|_{\infty}=1$, the inequality
\begin{equation*}
\|\lambda_1A_1+\dots+\lambda_NA_N\|_2\le \|H_1\|_2+\dots+\|H_N\|_2
\end{equation*}
holds. Thus,
\begin{equation}\label{disc1}
h_0\le \|H_1\|_2+\dots+\|H_N\|_2.
\end{equation}
We assume that there is a strict inequality in \eqref{disc1} and at least one of the coefficients $\eta_{kp}$ is not equal to zero. The result we presented in Theorem \ref{Thm2} will be better than the main result from \cite{BFCh2023} if inequality $M_{\eqref{5}}(\overline{\varepsilon})<M_{\eqref{Jul}}(\overline{\varepsilon})$ is satisfied. In this case, $\overline \varepsilon\in(0,\overline{\varepsilon}_{\max})$, where $\overline{\varepsilon}_{\max}$ is the only positive root of the equation
\begin{equation*}
\overline{\varepsilon}_{\max}\sum\limits_{k=2}^{N}\sum\limits_{p=1}^{\infty}\frac{\eta_{kp}(\overline{\varepsilon}_{\max})^{p-1}}{(p+1)!}=
\|H_1\|_2+\dots+\|H_N\|_2-h_0.
\end{equation*}
 If the Hamiltonians $H_k$, $k=1,\dots,N$ form a commuting family, then all constants $\eta_{kp}=0$. In this case, the result of Theorem \ref{Thm2} is always better than the main result from \cite{BFCh2023}, regardless of $\overline{\varepsilon}$. It is intuitively clear that for families of Hamiltonians close to commuting ones, our proposed approach leads to less conservative estimates of the errors of coherent states.
 
 This is the main conclusion of our work. The practical significance of this conclusion for the design of quantum computing devices is an open problem requiring further research.

 In the case when there is an equality in \eqref{disc1}, we do not arrive at a better result than in \cite{BFCh2023}.
  
It is also of interest for further research to expand our results for incoherent perturbations of a quantum mechanical system.

\section{Appendix A}
 Using the method of mathematical induction on the natural parameter $k$ and the commutator formula \eqref{0}, we prove the formula
\begin{equation}\label{AppA}
\gathered
e^{-i(1+\varepsilon_1 t) H_1}e^{-i(1+\varepsilon_2 t) H_2}\dots
e^{-i(1+\varepsilon_{k-1} t) H_{k-1}}(H_{k})\\
=e^{-i(1+\varepsilon_1 t)\ad H_1}e^{-i(1+\varepsilon_2t)\ad H_2}\dots e^{-i(1+\varepsilon_{k-1}t)\ad H_{k-1}}(H_{k})\\
\times e^{-i(1+\varepsilon_1 t)H_1}\dots e^{-i(1+\varepsilon_{k-1} t)H_{k-1}}
\endgathered
\end{equation}
The induction base follows directly from \eqref{0}. We assume that \eqref{AppA} is valid for some $k\ge 1$ and show that this formula is valid for $k+1$. Indeed,
\begin{equation*}
\gathered
e^{-i(1+\varepsilon_1 t) H_1}e^{-i(1+\varepsilon_2 t) H_2}\dots
e^{-i(1+\varepsilon_k t) H_k}(H_{k+1})\\
=e^{-i(1+\varepsilon_1 t) H_1}\times\big(e^{-i(1+\varepsilon_2t)\ad H_2}\dots e^{-i(1+\varepsilon_kt)\ad H_k}(H_{k+1})\big)\\
\times e^{-i(1+\varepsilon_2 t)H_2}\dots e^{-i(1+\varepsilon_k t)H_k}\\
=e^{-i(1+\varepsilon_1 t)\ad H_1}\big(e^{-i(1+\varepsilon_2t)\ad H_2}\dots e^{-i(1+\varepsilon_kt)\ad H_k}(H_{k+1})\big)\\
\times e^{-i(1+\varepsilon_1 t)H_1}\dots e^{-i(1+\varepsilon_k t)H_k}\\
=e^{-i(1+\varepsilon_1 t)\ad H_1}e^{-i(1+\varepsilon_2t)\ad H_2}\dots e^{-i(1+\varepsilon_kt)\ad H_k}(H_{k+1})\\
\times e^{-i(1+\varepsilon_1 t)H_1}\dots e^{-i(1+\varepsilon_k t)H_k},
\endgathered
\end{equation*}
which completes the proof \eqref{AppA}.

\section{Appendix B}
Using the method of mathematical induction on the natural parameter $k$, we prove the inequality
\begin{equation}\label{AppB}
\gathered
\|e^{-i(1+\varepsilon_1 t)\ad H_1}\dots
e^{-i(1+\varepsilon_{k-1} t)\ad H_{k-1}}(H_{k}) \\- 
e^{-i\ad H_1}\dots
e^{-i\ad H_{k-1}}(H_{k})\|_2\\
\le\sum\limits_{p=1}^{\infty}\frac{(\overline{\varepsilon}|t|)^p}{p!}
\Big(\|(\ad H_{k-1})^p(H_k)\|_2 \\ +\|(\ad H_{k-2})^p(e^{-i\ad H_{k-1}}(H_k))\|_2
+\dots\\
+\|(\ad H_{1})^p(e^{-i\ad H_2}\dots e^{-i\ad H_{k-1}}(H_k))\|_2\Big)
\endgathered
\end{equation}
The induction base for $k=2$ follows from \eqref{fact0}
\begin{equation}\label{AppB*}
\gathered
\|e^{-i(1+\varepsilon_1 t)\ad H_1}(H_2)-e^{-i\ad H_1}(H_2)\|_2\\
=\|e^{-i\ad H_1}(e^{-i\varepsilon_1t\ad H_1}(H_2)-H_2)\|_2 \\= 
\|e^{-it\varepsilon_1\ad H_1}(H_2)-H_2\|_2\\
\le
\sum\limits_{p=1}^{\infty}\frac{(\overline{\varepsilon} |t|)^p}{p!}\|(\ad H_1)^p(H_2)\|_2.
\endgathered
\end{equation}
We assume that \eqref{AppB} is true for $k\ge 2$ and derive \eqref{AppB} for $k+1$. Indeed, using the triangle inequality for the norm and \eqref{fact0}, we get
\begin{equation*}
\gathered
\|e^{-i(1+\varepsilon_1 t)\ad H_1}\dots
e^{-i(1+\varepsilon_{k-1} t)\ad H_{k-1}} \\
\times e^{-i(1+\varepsilon_{k} t)\ad H_{k}}(H_{k+1})
-e^{-i\ad H_1}\dots
e^{-i\ad H_{k-1}}\\ \times e^{-i\ad H_{k}}(H_{k+1})\|_2
\le\\
\|e^{-i(1+\varepsilon_1 t)\ad H_1}\dots
e^{-i(1+\varepsilon_{k-1} t)\ad H_{k-1}}\times\\e^{-i(1+\varepsilon_{k} t)\ad H_{k}}(H_{k+1})\\
-e^{-i(1+\varepsilon_1 t)\ad H_1}\dots
e^{-i(1+\varepsilon_{k-1} t)\ad H_{k-1}}e^{-i\ad H_{k}}(H_{k+1})\|_2\\
+\|e^{-i(1+\varepsilon_1 t)\ad H_1}\dots
e^{-i(1+\varepsilon_{k-1} t)\ad H_{k-1}}e^{-i\ad H_{k}}(H_{k+1})\\-
e^{-i\ad H_1}\dots
e^{-i\ad H_{k-1}}e^{-i\ad H_{k}}(H_{k+1})\|_2\\
\le
\|e^{-i(1+\varepsilon_{k} t)\ad H_{k}}(H_{k+1})-e^{-i\ad H_{k}}(H_{k+1})\|_2\\
+\sum\limits_{p=1}^{\infty}\frac{(\overline{\varepsilon}|t|)^p)}{p!}
\Big(\|(\ad H_{k-1})^p(e^{-i\ad H_k}(H_{k+1}))\|_2 \\+ 
\|(\ad H_{k-2})^p(e^{-i\ad H_{k-1}}e^{-i\ad H_k}(H_{k+1})\|_2\\
+\dots+\|(\ad H_1)^p(e^{-i\ad H_2}\dots e^{-i\ad H_k}(H_{k+1})\|_2\Big)\\
\le\sum\limits_{p=1}^{\infty}\frac{(\overline{\varepsilon}|t|)^p}{p!}
\Big(\|(\ad H_{k})^p(H_{k+1})\|_2\\+ \|(\ad H_{k-1})^p(e^{-i\ad H_{k}}(H_{k+1}))\|_2
+\dots\\
+\|(\ad H_{1})^p(e^{-i\ad H_2}\dots e^{-i\ad H_{k}}(H_{k+1}))\|_2\Big).
\endgathered
\end{equation*}

\section{Appendix C} 
We consider the problem of calculating the constant $h_0$ for the case of two matrices ($N=2$) of arbitrary dimension $n$. Let $A$, $B\in\Bbb C^{n\times n}$ matrices. We introduce the notation
\begin{equation*}
\gathered
\tau_0:=\tr(A^{\dagger} A),\quad \tau_1:=2\tr(A^{\dagger}B),\quad\tau_2:=\tr(B^{\dagger}B),\\
\mu_0:=\det A,\quad\mu_1:=
\begin{vmatrix}
a_{11}&b_{12}\\
a_{21}&b_{22}
\end{vmatrix}
+
\begin{vmatrix}
b_{11}&a_{12}\\
b_{21}&a_{22}
\end{vmatrix},
\quad
\mu_2:=\det B.
\endgathered
\end{equation*}
\begin{equation*}
\gathered
\omega_0:=|\mu_0|^2,\quad \omega_1:=2\Ree(\overline{\mu}_0\mu_1),
\quad \omega_2:=|\mu_1|^2+2\Ree(\overline{\mu}_0\mu_2),\\
\quad \omega_3:=2\Ree(\overline{\mu}_1\mu_2),\quad\omega_4:=|\mu_2|^2.
\endgathered
\end{equation*}

\begin{equation}
    \begin{split}
\zeta_0&=-\omega_{1}\tau_{0}\tau_{1}+\omega_{0}\tau_{1}^{2}+\omega_{1}^{2},\\        \zeta_1&=-2\tau_{0}\tau_{1}\omega_{2}-2\tau_{0}\omega_{1}\tau_{2}+4\omega_{0}\tau_{1}\tau_{2}+ 4\omega_{1}\omega_{2}, \\
        \zeta_2&=-3\tau_{0}\tau_{1}\omega_{3}-4\tau_{0}\tau_{2}\omega_{2}-\omega_{2}\tau_{1}^{2}
        +\omega_{1}\tau_{1}\tau_{2}+4\omega_{0}\tau_{2}^{2}\\ &+6\omega_{1}\omega_{3}+4\omega_{2}^{2},
 \\    \zeta_3&=-4\tau_{0}\tau_{1}\omega_{4}-6\tau_{0}\tau_{2}\omega_{3}-2\omega_{3}\tau_{1}^{2}-2
\omega_{2}\tau_{1}\tau_{2}+2\omega_{1}\tau_{2}^{2}\\ &+8\omega_{1}\omega_{4}+12\omega_{2}
\,\omega_{3},
 \\
        \zeta_4&=-8\tau_{0}\tau_{2}\omega_{4}-3\omega_{4}\tau_{1}^{2}-5\omega_{3}\tau_{1}\tau_{2}+
16\omega_{2}\omega_{4}+9\omega_{3}^{2},
 \\
        \zeta_5&=-8\tau_{1}\tau_{2}\omega_{4}-2\omega_{3}\,\tau_{2}^{2}+24\omega_{3}\omega_{4}, \\
        \zeta_6&=-4\tau_{2}^{2}\omega_{4}+16\omega_{4}^{2}. 
    \end{split}
\end{equation}
\begin{equation*}
\gathered
S:=\Big\{x\in(-1,1)\,|\,\sum\limits_{k=0}^6\zeta_kx^k=0\Big\}.
\endgathered
\end{equation*}
\begin{equation}\label{AppC*}
\gathered
\mathcal C:=\Big\{x\in(-1,1)\,|\, (\tau^2_0-4\omega_0)+(2\tau_0\tau_1-4\omega_1)x\\
+(\tau_1^2+2\tau_0\tau_2-4\omega_2)x^2+
(2\tau_1\tau_2-4\omega_3)x^3+(\tau_2^2-4\omega_4)x^4=0\Big\}.
\endgathered
\end{equation}
Let 
\begin{equation*}
\gathered
\Gamma(A,B):=\max\limits_{x\in S}\sqrt{\frac{\omega_1+2\omega_2x+3\omega_3x^2+4\omega_4x^3}{\tau_1+2\tau_2x}}.
\endgathered
\end{equation*}
{\bf Proposition.} Let
\begin{equation}\label{AppC0}
\gathered
2\omega_1\tau_2^3-2\tau_1\tau_2^2\omega_2+\frac{3}{2}\tau_2\omega_3\tau_1^2-\omega_4\tau_1^3\ne 0,
\endgathered
\end{equation}
and $\mathcal C=\emptyset$.

Then,
\begin{equation}\label{AppC1}
\gathered
h_0=\max_{(\lambda_1,\lambda_2)\in\Bbb R^2,\|\lambda\|_{\infty}=1}\|\lambda_1A+\lambda_2B\|_2\\=
\max\{\|A+B\|_2,\|A-B\|_2,\Gamma(A,B),\Gamma(B,A)\}.
\endgathered
\end{equation}
{\it Proof.} Let $x\in(-1,1)$, $\kappa(x)$ be an eigenvalue of matrix $(A+xB)^{\dagger}(A+xB)$. Then, $k(x)\ge 0$ and
% \begin{equation}
\begin{multline}\label{AppC1*}
    (\kappa(x))^2-\tr((A+xB)^{\dagger}(A+xB))\kappa(x) \\ +|\det(A+xB)|^2=0.
\end{multline}
% \end{equation}
 By direct calculations, we obtain
\begin{equation*}
\gathered
\tr((A+xB)^{\dagger}(A+xB))=\tau_0+\tau_1x+\tau_2x^2:=s(x),\\
|\det(A+x B)|^2=\omega_0+\omega_1x+\omega_2x^2+\omega_3x^3+\omega_4x^4:=d(x).
\endgathered
\end{equation*}
The $\mathcal C=\emptyset$ condition guarantees the possibility of applying the implicit function theorem, which implies that
\begin{equation*}
\gathered
2\kappa(x)\kappa^{\prime}(x)-s^{\prime}(x)\kappa(x)-s(x)\kappa^{\prime}(x)+d^{\prime}(x)=0.
\endgathered
\end{equation*}
If the function $\kappa(x)$ at the point $x_0\in(-1,1)$ takes an extremal value, then by Fermat's theorem $\kappa^{\prime}(x_0)=0$; therefore,
\begin{equation}\label{AppC2}
\gathered
-s^{\prime}(x_0)\kappa(x_0)+d^{\prime}(x_0)=0.
\endgathered
\end{equation}
The \eqref{AppC0} condition guarantees that $s^{\prime}(x_0)\ne 0$. Expressing from \eqref{AppC2} $\kappa(x_0)$ and substituting this expression into \eqref{AppC1*}, we find that
\begin{equation*}
\gathered
(d^{\prime}(x_0))^2-
s^{\prime}(x_0)s(x_0)d^{\prime}(x_0)+d(x_0)(s^{\prime}(x_0))^
2\\=\sum\limits_{k=0}^6\zeta_kx_0^k=0
\endgathered
\end{equation*}
Consequently, $S$ is the set of critical points of the function $\kappa(x)$ on the interval $(-1,1)$; therefore,
\begin{equation*}
\gathered
\max\limits_{x\in[-1,1]}\|A+xB\|_2=
\max\limits_{x_0\in S}\{\|A+B\|_2,\|A-B\|_2,\sqrt{\kappa(x_0)}
\}\\=
\max
\{\|A+B\|_2,\|A-B\|_2,\Gamma(A,B)\}.
\endgathered
\end{equation*}
Thus,
\begin{equation*}
\gathered
h_0=\max\limits_{\lambda\in\Bbb R^2,\|\lambda\|_{\infty}=1}\|\lambda_1A+\lambda_2B\|_2= \\
\max\{\max\limits_{x\in[-1,1]}{\|A+Bx\|_2},\max\limits_{x\in[-1,1]}{\|Ax+B\|_2}\}\\
=\max\{\|A+B\|_2,\|A-B\|_2,\Gamma(A,B),\Gamma(B,A)\}.
\endgathered
\end{equation*}
The proposition is proved.

\bibliographystyle{cas-model2-names}
\bibliography{mybibliob.bib}

\begin{thebibliography}{15}
\expandafter\ifx\csname natexlab\endcsname\relax\def\natexlab#1{#1}\fi
\providecommand{\url}[1]{\texttt{#1}}
\providecommand{\href}[2]{#2}
\providecommand{\path}[1]{#1}
\providecommand{\DOIprefix}{doi:}
\providecommand{\ArXivprefix}{arXiv:}
\providecommand{\URLprefix}{URL: }
\providecommand{\Pubmedprefix}{pmid:}
\providecommand{\doi}[1]{\href{http://dx.doi.org/#1}{\path{#1}}}
\providecommand{\Pubmed}[1]{\href{pmid:#1}{\path{#1}}}
\providecommand{\bibinfo}[2]{#2}
\ifx\xfnm\relax \def\xfnm[#1]{\unskip,\space#1}\fi
%Type = Article
\bibitem[{Atamas et~al.(2023)Atamas, Dashkovskiy and Slynko}]{ADSl2023}
\bibinfo{author}{Atamas, I.}, \bibinfo{author}{Dashkovskiy, S.}, \bibinfo{author}{Slynko, V.}, \bibinfo{year}{2023}.
\newblock \bibinfo{title}{Lyapunov functions for linear hyperbolic systems}.
\newblock \bibinfo{journal}{IEEE Transactions on Automatic Control} , \bibinfo{pages}{1--13}\DOIprefix\doi{10.1109/TAC.2023.3247879}.
%Type = Article
\bibitem[{Berberich et~al.(2023)Berberich, Fink and Holm}]{BFCh2023}
\bibinfo{author}{Berberich, J.}, \bibinfo{author}{Fink, D.}, \bibinfo{author}{Holm, C.}, \bibinfo{year}{2023}.
\newblock \bibinfo{title}{Robustness of quantum algorithms against coherent control errors}.
\newblock \bibinfo{journal}{arXiv preprint arXiv:2303.00618} .
%Type = Article
\bibitem[{Grushkovskaya et~al.(2018)Grushkovskaya, Zuyev and Ebenbauer}]{Grush}
\bibinfo{author}{Grushkovskaya, V.}, \bibinfo{author}{Zuyev, A.}, \bibinfo{author}{Ebenbauer, C.}, \bibinfo{year}{2018}.
\newblock \bibinfo{title}{On a class of generating vector fields for the extremum seeking problem: {L}ie bracket approximation and stability properties}.
\newblock \bibinfo{journal}{Automatica J. IFAC} \bibinfo{volume}{94}, \bibinfo{pages}{151--160}.
\newblock \URLprefix \url{https://doi.org/10.1016/j.automatica.2018.04.024}, \DOIprefix\doi{10.1016/j.automatica.2018.04.024}.
%Type = Article
\bibitem[{Jones(2003a)}]{8}
\bibinfo{author}{Jones, J.A.}, \bibinfo{year}{2003}a.
\newblock \bibinfo{title}{Robust ising gates for practical quantum computation}.
\newblock \bibinfo{journal}{Physical Review A} \bibinfo{volume}{67}, \bibinfo{pages}{012317}.
%Type = Article
\bibitem[{Jones(2003b)}]{jones}
\bibinfo{author}{Jones, J.A.}, \bibinfo{year}{2003}b.
\newblock \bibinfo{title}{Robust ising gates for practical quantum computation}.
\newblock \bibinfo{journal}{Phys. Rev. A} \bibinfo{volume}{67}, \bibinfo{pages}{012317}.
\newblock \URLprefix \url{https://link.aps.org/doi/10.1103/PhysRevA.67.012317}, \DOIprefix\doi{10.1103/PhysRevA.67.012317}.
%Type = Article
\bibitem[{Levitt(1986a)}]{7}
\bibinfo{author}{Levitt, M.H.}, \bibinfo{year}{1986}a.
\newblock \bibinfo{title}{Composite pulses}.
\newblock \bibinfo{journal}{Progress in Nuclear Magnetic Resonance Spectroscopy} \bibinfo{volume}{18}, \bibinfo{pages}{61--122}.
%Type = Article
\bibitem[{Levitt(1986b)}]{LEVITT}
\bibinfo{author}{Levitt, M.H.}, \bibinfo{year}{1986}b.
\newblock \bibinfo{title}{Composite pulses}.
\newblock \bibinfo{journal}{Progress in Nuclear Magnetic Resonance Spectroscopy} \bibinfo{volume}{18}, \bibinfo{pages}{61--122}.
\newblock \URLprefix \url{https://www.sciencedirect.com/science/article/pii/007965658680005X}, \DOIprefix\doi{https://doi.org/10.1016/0079-6565(86)80005-X}.
%Type = Book
\bibitem[{Magnus et~al.(2004)Magnus, Karrass and Solitar}]{MKS}
\bibinfo{author}{Magnus, W.}, \bibinfo{author}{Karrass, A.}, \bibinfo{author}{Solitar, D.}, \bibinfo{year}{2004}.
\newblock \bibinfo{title}{Combinatorial Group Theory: Presentations of Groups in Terms of Generators and Relations}.
\newblock Dover books on mathematics, \bibinfo{publisher}{Dover Publications}.
\newblock \URLprefix \url{https://books.google.com/books?id=1LW4s1RDRHQC}.
%Type = Inbook
\bibitem[{Merrill and Brown(2014a)}]{9}
\bibinfo{author}{Merrill, J.T.}, \bibinfo{author}{Brown, K.R.}, \bibinfo{year}{2014}a.
\newblock \bibinfo{title}{Progress in Compensating Pulse Sequences for Quantum Computation}. \bibinfo{publisher}{John Wiley \& Sons, Ltd}.
\newblock pp. \bibinfo{pages}{241--294}.
\newblock \URLprefix \url{https://onlinelibrary.wiley.com/doi/abs/10.1002/9781118742631.ch10}, \DOIprefix\doi{https://doi.org/10.1002/9781118742631.ch10}, \href{http://arxiv.org/abs/https://onlinelibrary.wiley.com/doi/pdf/10.1002/9781118742631.ch10}{\tt arXiv:https://onlinelibrary.wiley.com/doi/pdf/10.1002/9781118742631.ch10}.
%Type = Inbook
\bibitem[{Merrill and Brown(2014b)}]{merill}
\bibinfo{author}{Merrill, J.T.}, \bibinfo{author}{Brown, K.R.}, \bibinfo{year}{2014}b.
\newblock \bibinfo{title}{Progress in Compensating Pulse Sequences for Quantum Computation}. \bibinfo{publisher}{John Wiley \& Sons, Ltd}.
\newblock pp. \bibinfo{pages}{241--294}.
\newblock \URLprefix \url{https://onlinelibrary.wiley.com/doi/abs/10.1002/9781118742631.ch10}, \DOIprefix\doi{https://doi.org/10.1002/9781118742631.ch10}, \href{http://arxiv.org/abs/https://onlinelibrary.wiley.com/doi/pdf/10.1002/9781118742631.ch10}{\tt arXiv:https://onlinelibrary.wiley.com/doi/pdf/10.1002/9781118742631.ch10}.
%Type = Article
\bibitem[{Skolik et~al.(2023)Skolik, Mangini, B{\"a}ck, Macchiavello and Dunjko}]{28}
\bibinfo{author}{Skolik, A.}, \bibinfo{author}{Mangini, S.}, \bibinfo{author}{B{\"a}ck, T.}, \bibinfo{author}{Macchiavello, C.}, \bibinfo{author}{Dunjko, V.}, \bibinfo{year}{2023}.
\newblock \bibinfo{title}{Robustness of quantum reinforcement learning under hardware errors}.
\newblock \bibinfo{journal}{EPJ Quantum Technology} \bibinfo{volume}{10}, \bibinfo{pages}{1--43}.
%Type = Article
\bibitem[{Slyn'ko et~al.(2019)Slyn'ko, Tun\c{c} and Bivziuk}]{BSlT2019}
\bibinfo{author}{Slyn'ko, V.I.}, \bibinfo{author}{Tun\c{c}, O.}, \bibinfo{author}{Bivziuk, V.O.}, \bibinfo{year}{2019}.
\newblock \bibinfo{title}{Application of commutator calculus to the study of linear impulsive systems}.
\newblock \bibinfo{journal}{Systems Control Lett.} \bibinfo{volume}{123}, \bibinfo{pages}{160--165}.
\newblock \URLprefix \url{https://doi.org/10.1016/j.sysconle.2018.10.015}, \DOIprefix\doi{10.1016/j.sysconle.2018.10.015}.
%Type = Article
\bibitem[{Suttner(2023)}]{Sutner}
\bibinfo{author}{Suttner, R.}, \bibinfo{year}{2023}.
\newblock \bibinfo{title}{Nonsmooth optimization by {L}ie bracket approximations into random directions}.
\newblock \bibinfo{journal}{Systems Control Lett.} \bibinfo{volume}{174}, \bibinfo{pages}{Paper No. 105481, 8}.
\newblock \URLprefix \url{https://doi.org/10.1016/j.sysconle.2023.105481}, \DOIprefix\doi{10.1016/j.sysconle.2023.105481}.
%Type = Article
\bibitem[{Trout et~al.(2018)Trout, Li, Gutiérrez, Wu, Wang, Duan and Brown}]{5}
\bibinfo{author}{Trout, C.J.}, \bibinfo{author}{Li, M.}, \bibinfo{author}{Gutiérrez, M.}, \bibinfo{author}{Wu, Y.}, \bibinfo{author}{Wang, S.T.}, \bibinfo{author}{Duan, L.}, \bibinfo{author}{Brown, K.R.}, \bibinfo{year}{2018}.
\newblock \bibinfo{title}{Simulating the performance of a distance-3 surface code in a linear ion trap}.
\newblock \bibinfo{journal}{New Journal of Physics} \bibinfo{volume}{20}, \bibinfo{pages}{043038}.
\newblock \URLprefix \url{https://dx.doi.org/10.1088/1367-2630/aab341}, \DOIprefix\doi{10.1088/1367-2630/aab341}.
%Type = Article
\bibitem[{Zuyev(2016)}]{Zu}
\bibinfo{author}{Zuyev, A.}, \bibinfo{year}{2016}.
\newblock \bibinfo{title}{Exponential stabilization of nonholonomic systems by means of oscillating controls}.
\newblock \bibinfo{journal}{SIAM J. Control Optim.} \bibinfo{volume}{54}, \bibinfo{pages}{1678--1696}.
\newblock \URLprefix \url{https://doi.org/10.1137/140999955}, \DOIprefix\doi{10.1137/140999955}.

\end{thebibliography}

\end{document}